\documentclass[conference]{IEEEtran}
\IEEEoverridecommandlockouts
\usepackage{cite}
\usepackage{amsmath,amssymb,amsfonts}
\usepackage{algorithmic}
\usepackage{graphicx}
\usepackage{textcomp}
\usepackage{xcolor}
\usepackage{booktabs}
\usepackage{caption}
\usepackage[table]{xcolor}
\usepackage{makecell}
\usepackage[ruled,vlined]{algorithm2e}

\def\BibTeX{{\rm B\kern-.05em{\sc i\kern-.025em b}\kern-.08em
    T\kern-.1667em\lower.7ex\hbox{E}\kern-.125emX}}
\begin{document}

\title{Experimental Multi-site Testbed for Advanced Control and Optimization of Hybrid Energy Systems\\

\thanks{This material is based upon work supported by the U.S. Department of Energy’s Office of EERE under award number DE-EE0010147. The views expressed herein do not necessarily represent the views of the U.S. Department of Energy or the United States Government.}
}

\author{\IEEEauthorblockN{Arash Omidi, Tanmay Mishra, Mads R. Almassalkhi} \IEEEauthorblockA{\textit{Department of Electrical and Biomedical Engineering} \\ \textit{University of Vermont}\\ Burlington, Vermont, USA \\ \{aomidi, tmishra, malmassa\}@uvm.edu}  }

\maketitle

\begin{abstract}
This paper presents a hybrid energy system (HES) experimental testbed developed at the University of Vermont, featuring a dual-site architecture that integrates on-campus laboratory facility with an off-campus solar and meteorological station. This supports the prototyping and validation of advanced HES control and optimization strategies. The platform integrates hardware-in-the-loop (HIL) simulations with a reconfigurable set of kVA-scale assets. 
A unified monitoring and communication architecture supports real-time data acquisition, model validation, and control implementation. The capabilities of the testbed are demonstrated through an HIL experiment in which a battery systems participate in solar PV smoothing.
\end{abstract}

\begin{IEEEkeywords}
Experimental testbed, hardware-in-the-loop validation, hybrid energy systems, PV smoothing.   
\end{IEEEkeywords}
\vspace{-5pt}

\section{Introduction}
The global transition toward sustainable energy has accelerated renewable energy deployment. However, the inherent variability and uncertainty of these resources challenges grid reliability, flexibility, and power quality~\cite{liang2016emerging}. Hybrid Energy Systems (HES) have emerged as a promising solution to address the growing global demand for reliable, efficient, and sustainable power. HES integrate diverse energy resources, such as intermittent renewable sources like photovoltaic (PV) and wind generation, energy storage, and controllable loads including electrolyzers, electric vehicle (EV) charging stations, or data centers to enhance flexibility and reliability compared to standalone systems~\cite{nassar2025role}. This has driven the rapid expansion of HES installations in the United States 
~\cite{Berkeley2025}.

Deploying HES at scale introduces new layers of operational and control complexity, including power converter interactions, energy and ramping constraints, coupled dynamics, and multi-timescale coordination~\cite{das2025research}. Most existing research is either limited to pairs of HES components (e.g., PV$ + X$) or relies on mathematical modeling and simulation. Consequently, these studies are primarily evaluated based on model accuracy, computational efficiency, and numerical solvers. Furthermore, these models often fail to capture nonlinear switching effects, communication delays, and hardware limits. Conversely, full-scale field testing is often infeasible due to cost, safety issues, and complexity. This creates a validation gap, highlighting the need for intermediate methodologies combining hardware testing fidelity with laboratory controllability and safety~\cite{stanev2020real}.


Laboratory-based testbeds bridge this validation gap by providing controlled and repeatable environments to study complex device interactions. Hardware-in-the-loop (HIL) simulation has become critical for these validation efforts. In an HIL configuration, physical hardware components are interfaced with real-time digital simulations to enable systematic evaluation of device and controller performance and system-level stability before field deployment. A multi-site HIL testbed refers to geographically separated facilities that exchange measurements and/or setpoints to support coordinated experiments. Such architectures can combine laboratory controllability with field realism, however they also introduce additional communication and integration requirements compared to single-site setups~\cite{essakiappan2021multi}. Several HIL testbeds have been developed that address different aspects of HES validation and inverter-dominated grid integration, as summarized in Table~\ref{tab:testbeds}.  While these facilities demonstrate valuable capabilities, most platforms emphasize either large-scale integration or single-facility component validation. Therefore, there remains a need for reconfigurable, intermediate-scale platforms that are multi-site and can incorporate field measurements into laboratory HIL studies, enabling multi-timescale HES control validation under real-world conditions.

\begin{table*}[!t]
\caption{Summary of selected North America testbeds and their key assets and capabilities}
\label{tab:testbeds}
\centering
\scriptsize
\begin{tabular}{@{}p{2.6cm}p{1.4cm}p{0.4cm}p{3.65cm}p{5cm}p{1.3cm}p{0.70cm}}
\toprule
\textbf{Testbed} &\textbf{Location} &\textbf{Scale} & \textbf{Key Assets} & \textbf{Capabilities} & \textbf{HES Integration} & \textbf{Multi-site}  \\
\midrule
Energy System Integration Facility (NREL)~\cite{kroposki2012energy} &Colorado& 1~MW& Power converters, batteries, electrolyzer, and fuel cell& HIL validation, power management evaluation, power system emulation, and microgrid control. & Yes&Yes\\
\hline
HIL Platform (North Carolina State University)~\cite{ghanbari2019hardware}&North Carolina&250~kW & PV and power converters & MPPT implementation, voltage and frequency control, grid emulation, and HIL validation. & No&No \\
\hline
FIU testbed (Florida International University)~\cite{youssef2015dds} &Florida&40~kW& AC generators, power converters, adjustable loads, and SCADA system & PV and wind emulation, HES AC/DC control, energy management, and MPPT implementation. & Limited&No\\
\hline
SSMTB testbed (Sandia)~\cite{glover2012secure}&New Mexico&30~kW& Diesel generators, energy storage, and programmable loads & PV and wind emulation, voltage and frequency control, and HIL validation. & Limited&No\\
\hline
UTA testbed (University of Texas-Arlington)~\cite{turner2014design}&Texas&15~kW& PV, wind turbine, fuel cell, battery, power converters, and loads & Power management, voltage and frequency control, and power quality analysis. & Yes &No\\
\hline
Multi-site networked HIL platform~\cite{essakiappan2021multi} & Three states (CO,NC,TN)& N/A &ADMS, protocol adapters, HIL sites & DER emulation, multi-site HIL evaluation of distributed control, and ADMS–DER coordination &No & Yes \\

\bottomrule
\end{tabular}
\vspace{-10pt}
\end{table*}

This paper presents a new multi-site HES experimental testbed at the University of Vermont (\emph{The Accelerated Testing Lab}), designed for the prototyping and validation of advanced control and coordination strategies for grid services. Compared to existing testbeds (Table~\ref{tab:testbeds}), the platform uniquely combines:
\begin{itemize}
    \item A dual-site architecture integrating off-site field PV and weather data with on-campus hardware assets and HIL,
    \item A reconfigurable HES platform with PV, batteries, inverters, and electrolyzer with plug-and-play interfaces,
    \item A unified framework for multi-timescale HES control and real-time data streaming.
\end{itemize}
This paper provides a comprehensive overview of the key capabilities and features of this testbed, and experimentally validates battery participation in solar smoothing as a representative test case. The validation is conducted through an HIL setup with a physical battery, simulated batteries, and a real-time controller. The remainder of this paper is organized as follows. Section~\ref{sec:testbed} describes the HES laboratory design and HIL configuration of the testbed, including the electrical system architecture and data streaming framework. Section~\ref{sec:Capabilities} details the key capabilities of the HES platform. Section~\ref{sec:test_case} presents an experimental test case, while Section~\ref{sec:conclusion} concludes the paper with perspectives on future work.

\section{Experimental testbed and system architecture}
\label{sec:testbed}

\begin{figure}[t]
\captionsetup{aboveskip=1pt,belowskip=-15pt} 
    \centering
    \includegraphics[width=0.39\textwidth]{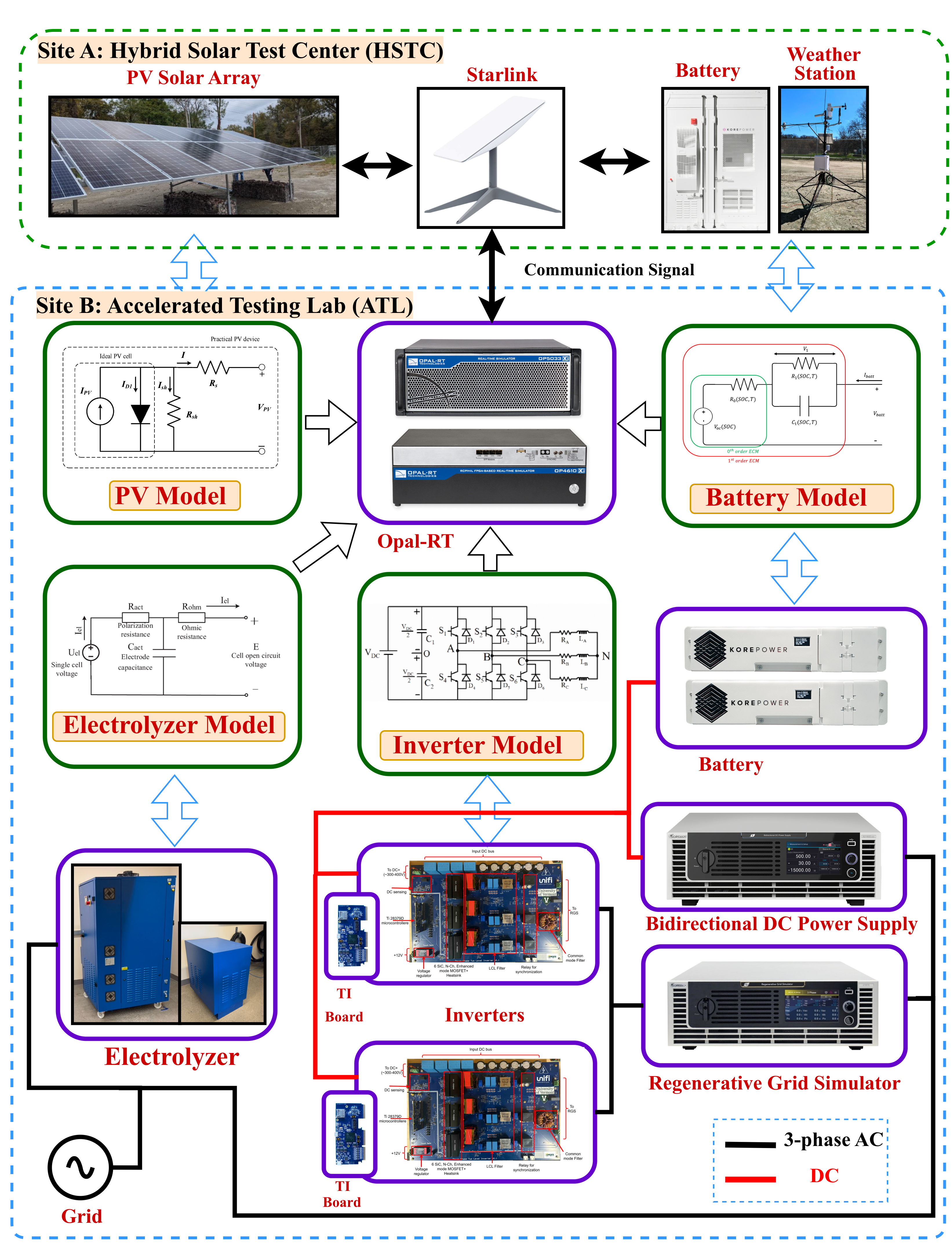}
    \caption{Multi-site HES testbed with plug-and-play architecture. Components simulated in OPAL-RT can be replaced with physical ATL hardware (inverters, batteries, electrolyzer) or real-time HSTC field data (PV, weather) streamed via Starlink (Fig.~\ref{Data_Streaming}), enabling flexible validation from pure simulation to HIL testing with real measurements.}
    \label{system_architecture}
\end{figure}

The HES testbed features a distributed architecture composed of two physically distinct but tightly interconnected laboratory facilities:
\begin{itemize}
\item The Accelerated Testing Laboratory (ATL), the primary on-campus HIL platform.
\item The Hybrid Solar Test Center (HSTC), an off-campus 23~kW solar and meteorological site with a planned battery system installation.
\end{itemize}
These facilities share real-time data, control signals, and HIL interfaces, enabling coordinated experiments that span devices, controllers, and grid-level scenarios. The overall system architecture is shown in Fig.~\ref{system_architecture}, with key asset specifications in Table~\ref{tab:equipment_specs}. The design and capabilities of each facility are detailed in the following subsections.

\begin{table}[t]
\caption{Equipment Specifications}
\label{tab:equipment_specs}
\centering
\scriptsize
\begin{tabular}{@{}lll@{}}
\toprule
\textbf{Component} & \textbf{Description} & \textbf{Specifications} \\
\midrule
PV Array & ET Solar & 23.4~kW DC\\
Inverter& SolarEdge & 3×10~kVA, single-phase\\
Battery (TBI 2026)& KORE Power& 100~kW/233~kWh\\
LFP Battery & Enphase ACB 1.0 & 6×1.2~kWh, 26.5~V \\
NMC Battery  & KORE Power & 2×6.5~kWh, 59.6~V \\
Inverter & UNIFI reference design & 4×5~kVA, 3-phase  \\
Electrolyzer & Verde hydrogen & 15~kW alkaline stack \\
DC Power Supply & Chroma 62120D-1200 & 12~kW, 0–1200~V, $\pm$55~A \\
Regenerative Grid Simulator & Chroma 61812 & 12~kVA, 0–600~V \\
Microcontroller & TMDSCNCD28379D & 100~MHz  \\
Real-Time Simulator & Opal-RT OP5033 & Intel® Xeon® (4 cores)\\
FPGA and I/O Expansion & Opal-RT OP4610 & FPGA, 64 A/D I/O\\
\bottomrule
\end{tabular}
\vspace{-20pt}
\end{table}
\vspace{-5pt}
\subsection{Accelerated Testing Laboratory}
The ATL at the University of Vermont (Fig.~\ref{system_architecture}) integrates hardware and simulation capabilities to enable comprehensive testing of power electronics, energy storage, and control strategies in a reconfigurable cyber-physical environment. A bidirectional DC power supply feeds either two parallel-connected 5~kVA inverters or the batteries. The inverters, designed based on the UNIFI reference design~\cite{UNIFI2025}, enable testing of advanced control strategies and the investigation of inverter dynamics and their impact on the grid. Each inverter is controlled by a Texas Instruments (TI) microcontroller, which enables the implementation of control strategies such as grid-forming (GFM) and grid-following (GFL) control, active/reactive power regulation, and coordinated multi-inverter operation. In addition, the DC power supply is connected to six 1.2~kWh Enphase LFP batteries, allowing for systematic charge, discharge, and storage co-optimization studies. On the AC side, the inverters are connected to a regenerative grid simulator (RGS). The testbed is coupled to a 15~kW electrolyzer, which represents hydrogen production as a controllable grid-interfaced load.

 Furthermore, the Opal-RT simulator, with the OP4610 I/O expansion unit, provides the computational environment to emulate detailed grid behaviors. As shown in Fig.~\ref{system_architecture}, the testbed architecture is explicitly designed for \emph{plug-and-play} flexibility. This allows any component initially simulated within the Opal-RT to be seamlessly replaced by its physical hardware counterpart for full HIL validation. 
 This creates a robust cyber-physical loop where Opal-RT issues analog setpoints to the RGS and bidirectional DC supply, which act as the physical interface to the device under test. These controllable power supplies generate the precise power and voltage conditions dictated by the real-time simulation. Voltage and current sensors capture responses from the hardware and feed them back to the Opal-RT's analog inputs, completing the HIL setup. This closed-loop configuration ensures that the real-time model dynamically reacts to physical hardware behavior and vice versa. This setup also facilitates integration of PV and weather profiles from the HSTC (discussed in the next subsection) to drive realistic validation scenarios.
\vspace{-5pt}
\subsection{Hybrid Solar Test Center}
The ATL is supported by the HSTC located at the McNeil generating plant, one mile from campus. The HSTC hosts a 23~kW grid-connected solar PV array with SolarEdge inverters and an on-site weather station that records solar irradiance, temperature, wind speed, and humidity every 5 seconds, providing synchronized environmental and electrical data for PV modeling and performance analysis. A battery energy storage system (BESS) will be installed to enable field validation of battery-solar coordination strategies and grid services under real-world conditions.
\vspace{-5pt}
\subsection{Multi-site Communication and Data Integrity}
\label{subsec:multisite_considerations}
\begin{figure}[t]
\captionsetup{aboveskip=1pt,belowskip=-15pt} 
    \centering    \includegraphics[width=0.42\textwidth]{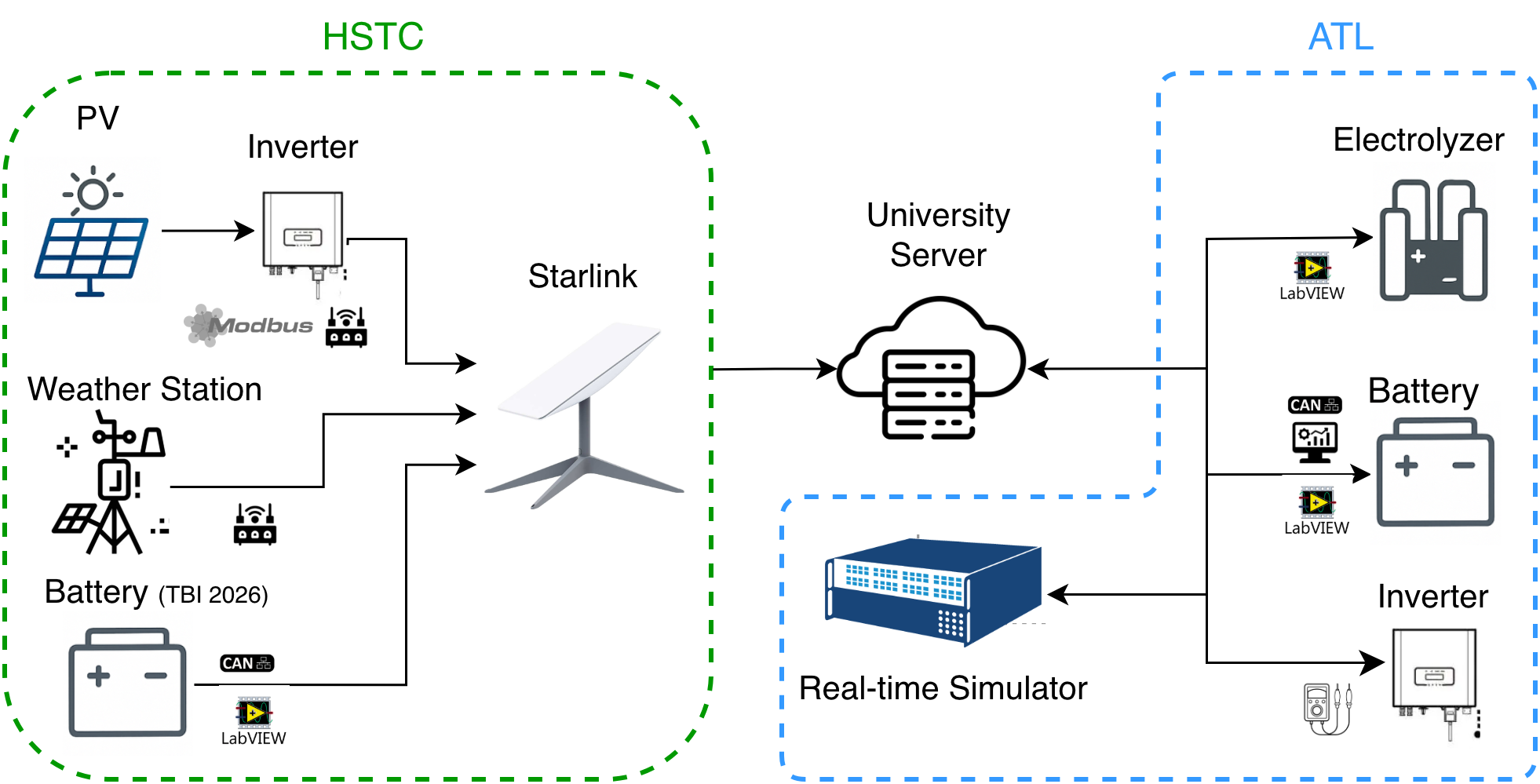}
    \caption{Data-communication architecture for HSTC and ATL.}
    \label{Data_Streaming}
\end{figure}

The ATL and HSTC facilities are interconnected through a communication architecture that employs a Starlink satellite network for real-time data streaming and an on-campus server. As shown in Fig.~\ref{Data_Streaming}, environmental and operational data from HSTC are measured using high-precision sensors and transmitted to a centralized server through Modbus coupled with a communication gateway. All data streams are time-synchronized and stored on the server for visualization and analysis, enabling real-time monitoring, control, and performance evaluation of HES subsystems.

While multi-site architectures offer significant advantages for HES research by integrating realistic field data with laboratory HIL capabilities, they also presents technical challenges that must be carefully managed. Communication latency between geographically separated sites can degrade closed-loop control performance, especially for fast-responding applications such as frequency regulation and voltage control, which require sub-second coordination. Precise time synchronization across sites is critical to maintain data integrity. Furthermore, integrating diverse components with different communication protocols (Modbus, CAN Bus) creates interoperability complexity. High-frequency measurements from distributed sensors also generate large data volumes that require robust infrastructure for time-aligned storage, real-time processing, and post-experiment analysis. The architecture in this paper addresses these challenges through a robust centralized server for high-volume data management, protocol-specific scripts with built-in time synchronization for each communication standard, and direct data streaming via Starlink to reduce communication delays in control loops.
\vspace{-5pt}

\section{Platform Capabilities and Features}
\label{sec:Capabilities}
The experimental testbed provides a comprehensive environment for investigating HES integration and advanced control validation. The platform key capabilities span three interconnected domains: 
\begin{itemize}
\item Component and system-level modeling and validation
\item Control implementation and verification
\item Hardware-in-the-loop (HIL) integration
\end{itemize}
This section describes the platform capabilities and their application to contemporary challenges in power system operation.

\subsection{Component and system-level modeling and validation}
Accurate models of individual HES components and their interactions are fundamental to enabling informed decision-making in system design, control development, and operational optimization. The ATL provides comprehensive experimental capabilities to validate physics-based and data-driven models across multiple components and timescales. The bidirectional DC power supply enables validation of battery models through controlled charge-discharge cycling, including the energy reservoir model~\cite{omidi2025towards} and electro-thermal equivalent circuit model~\cite{mishra2024predictive} developed in prior work. Physics-based electrolyzer models describing hydrogen production rate, efficiency, and thermal dynamics have been validated through power cycling experiments on the 15~kW alkaline stack~\cite{ayubirad2025modeling}. The platform also enables validation of efficiency-based and detailed switching inverter models. Data from the HSTC support PV model validation across diverse operating conditions and temperature dependencies. At the system level, aggregated circuit models capture interactions between PV, battery, and inverter systems through equivalent circuit modeling frameworks, facilitating state estimation and parameter identification~\cite{sang2024circuit}. These validated models form the foundation for multi-objective optimization and control studies.

\subsection{Control implementation and verification}

HES require coordinated controller design due to strong electrical and thermal coupling between components. As power systems become increasingly inverter dominated, traditional sources of inertia and voltage support disappear, and these services must be provided through tightly integrated, multi-layer control across HES components. Addressing these challenges requires experimental validation of advanced control and coordination strategies under realistic operating conditions. The ATL testbed bridges this gap by providing a flexible platform to validate control algorithms at component and system levels. At the component level, ATL enables direct implementation and testing of inverter and battery control algorithms, such as GFM and GFL modes, active/reactive power regulation, Volt-VAR control, and multi-inverter coordination, under programmable conditions including voltage sags, weak grid operation, and fault events. At the system level, the platform supports hierarchical and distributed control schemes, including model predictive control and economic dispatch-based coordination of batteries, PV, and electrolyzers. These validated control strategies enable the ATL to test algorithms for providing grid services, summarized in Table~\ref{tab:dev_methods}. The shaded grid service is experimentally validated in this paper.

\begin{table}[t]
\caption{Capabilities of Multi-site Testbed }
\label{tab:dev_methods}
\centering
\scriptsize
\setlength{\tabcolsep}{2pt}
\renewcommand{\arraystretch}{1.1}
\resizebox{\columnwidth}{!}{%
\begin{tabular}{@{}lcccc@{}}
\toprule
\textbf{Grid Services} &
\makecell{\textbf{Time scale}} &
\makecell{\textbf{Participating assets}} \\
\midrule
Synthetic damping/inertia        & $<$1s & Battery+Inverter  \\
Voltage regulation     & $<$60s & Battery/PV + Inverter  \\
Frequency regulation         &  seconds-minutes & Battery/Coupled HES   \\
\rowcolor{green!20}
Solar smoothing & seconds-minutes & PV+ Battery \\
Energy arbitrage \& peak shaving & minutes-hours& Coupled HES \\
\bottomrule
\end{tabular}%
}\vspace{-15pt}
\end{table}

\subsection{Hardware-in-the-loop (HIL) integration}
HIL validation bridges the gap between simulation and field deployment by integrating real hardware with real-time digital simulation. It enables realistic evaluation of controllers, converters, and system interactions while maintaining safety and flexibility. The ATL testbed, equipped with an OPAL-RT real-time simulator and I/O unit, supports HIL configurations. The OPAL-RT simulator supports steady-state and dynamic grid models, enabling comprehensive evaluation of hardware interactions with the grid across different timescales. This enables detailed studies of inverter–grid interactions, coordination among HES components, and validation of advanced control strategies under realistic operating scenarios.

\section{Experimental Test Case: Solar Smoothing using Physical Battery Storage}
\label{sec:test_case}

The increasing penetration of PV generation introduces rapid power fluctuations from cloud transitions and irradiance variability, which challenges grid stability and power quality. A BESS can mitigate these fluctuations through fast response and bidirectional power capability, absorbing excess power during PV spikes and injecting power during drops. By coordinating PV generation with BESS operation, HES can smooth the power delivered to the grid. Grid operators impose ramp rate limits to keep fluctuations within acceptable ranges, defined as:
\begin{equation}
\text{Ramp Rate: } RR(t) = \frac{P_\text{PV}(t) - P_\text{PV}(t-\Delta t)}{\Delta t \cdot P_{\text{rated}}} \times 100
\end{equation}
where $P_\text{PV}(t)$ is the PV power at time $t$, $\Delta t$ is the time interval, and $P_{\text{rated}}$ is the rated PV capacity. In this study, the maximum allowable ramp rate is set to $\pm10~\%$/min, consistent with limits specified in PV grid codes~\cite{martins2019comparative}.

\begin{figure}[t]
\captionsetup{aboveskip=1pt,belowskip=-15pt} 
    \centering
    \includegraphics[width=0.75\linewidth]{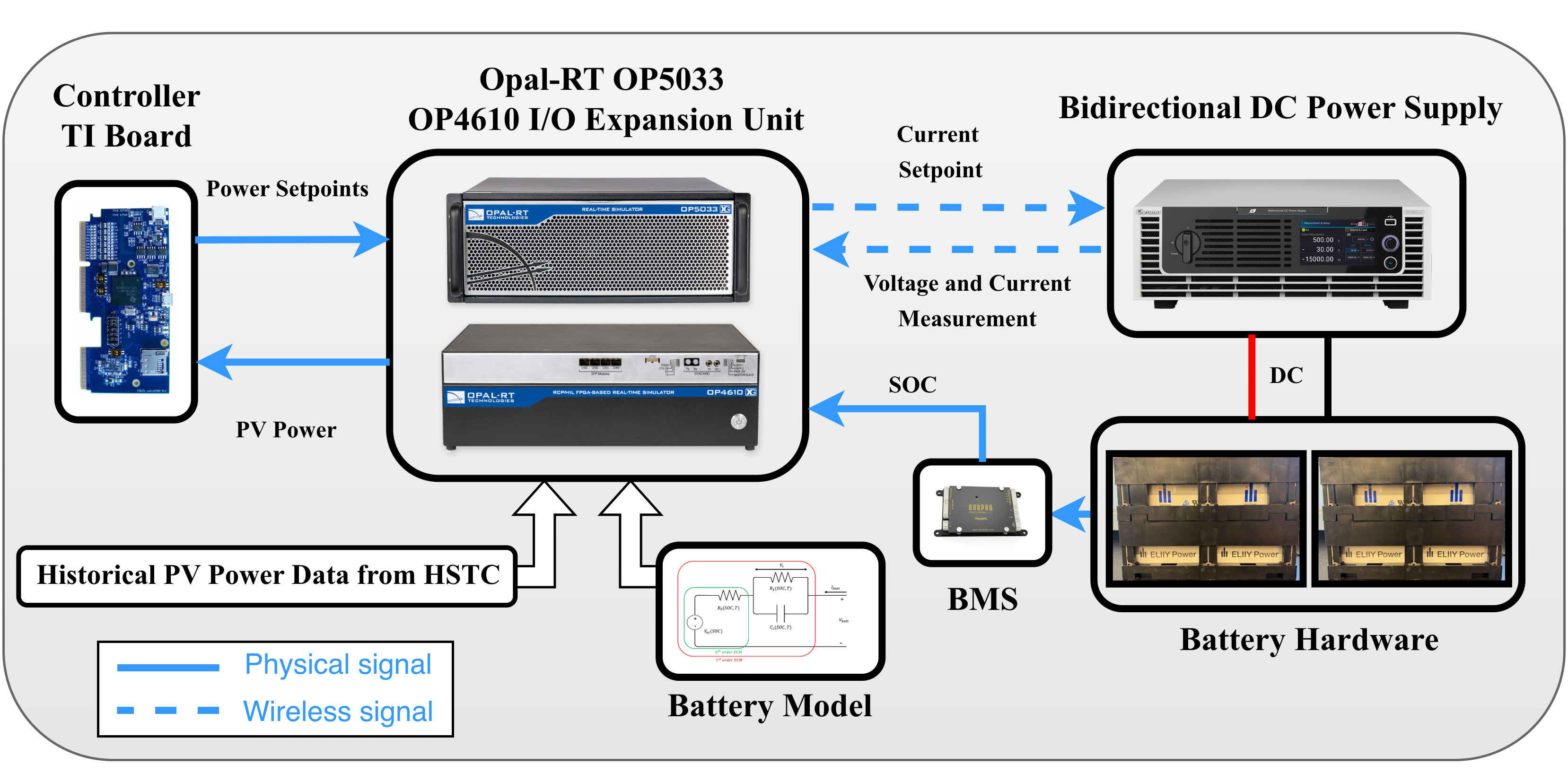}
    \caption{Experimental setup for solar smoothing HIL validation showing two series-connected Enphase batteries with a BMS, bidirectional DC power supply, TI control card, and OPAL-RT real-time simulator.}
    \label{fig:HIL_SM}
\end{figure}



\begin{algorithm}[b]
\scriptsize
\DontPrintSemicolon
\RestyleAlgo{plain}
\setlength{\algomargin}{0.6em}
\setlength{\interspacetitleruled}{1pt}
\setlength{\interspacealgoruled}{1pt}
\setlength{\belowcaptionskip}{1pt}
\setlength{\abovecaptionskip}{1pt}
\renewcommand{\baselinestretch}{0.6}\selectfont
\SetAlgoSkip{2pt}
\newcommand{\MyCommentStyle}[1]{\scriptsize\itshape #1}
\SetCommentSty{MyCommentStyle}
\SetSideCommentRight
\caption{PV Smoothing Controller}
\label{SmoothingAlgorithm}
\SetKwInOut{Input}{Input}\SetKwInOut{Output}{Output}

\Input{$P_{\mathrm{PV}}[k]$}
\BlankLine
\Output{$P_{\mathrm{batt,h}}[k],\ P_{\mathrm{batt,s}}[k]$}

\BlankLine
\textbf{Parameters:}
$N,\ T_s$, $w_h \gets \frac{P_{h}^{\max}}{P_{h}^{\max} + P_{s}^{\max}}$;\ $w_s \gets 1 - w_h$\;

\BlankLine
$P_{\mathrm{buf}}[i] \gets 0,\ i=1{:}N$;\ $k \gets 1$\tcp*{Initialization}\;

\While{controller is running}{
\BlankLine

  $P_{\mathrm{buf}}[([k-1]\bmod N)+1] \gets P_{\mathrm{PV}}[k]$
  \tcp*{1) Insert new PV sample}\;

  $\hat{P}_{\mathrm{PV}}[k] \gets \frac{1}{N}\sum_{i=1}^N P_{\mathrm{buf}}[i]$
  \tcp*{2) Smoothed PV power}\;

  $P_{\mathrm{batt}}[k] \gets P_{\mathrm{PV}}[k] - \hat{P}_{\mathrm{PV}}[k]$
  \tcp*{3) Total battery command}\;
  
  $P_{\mathrm{batt,h}}[k] \gets w_h \cdot P_{\mathrm{batt}}[k]$ \tcp*{4) Battery power allocation}\;
  $P_{\mathrm{batt,s}}[k] \gets w_s \cdot P_{\mathrm{batt}}[k]$\;

  \BlankLine
  Output $P_{\mathrm{batt,h}}[k],\ P_{\mathrm{batt,s}}[k]$ \tcp*{5) Output}\;
  $k \gets k+1$, sleep($T_s$) \;
}
\end{algorithm}

To evaluate the smoothing capability at ATL, an HIL experiment was conducted using the experimental setup shown in Fig.~\ref{fig:HIL_SM}. The experiment uses historical 5-second PV power data from the HSTC site. To meet the power requirements for PV smoothing, the testbed integrates physical battery hardware (1.2~kW/2.4~kWh) together with three battery equivalent circuit models connected in series and implemented as the simulated battery (7.2~kW/7.2~kWh). The OPAL-RT sends PV power to the TI controller, which runs a 30-minute moving-average smoothing algorithm, detailed in Algorithm~\ref{SmoothingAlgorithm}. An energy management system (EMS) distributes the required battery power between the two battery systems based on their maximum power ratings. The controller issues setpoints for the battery to the power supply through OPAL-RT. The battery management system (BMS) estimates the battery SOC.

\begin{figure}[t]
\captionsetup{aboveskip=1pt,belowskip=-15pt} 
    \centering

    \begin{minipage}{0.45\textwidth}
        \centering
        \includegraphics[width=\linewidth]{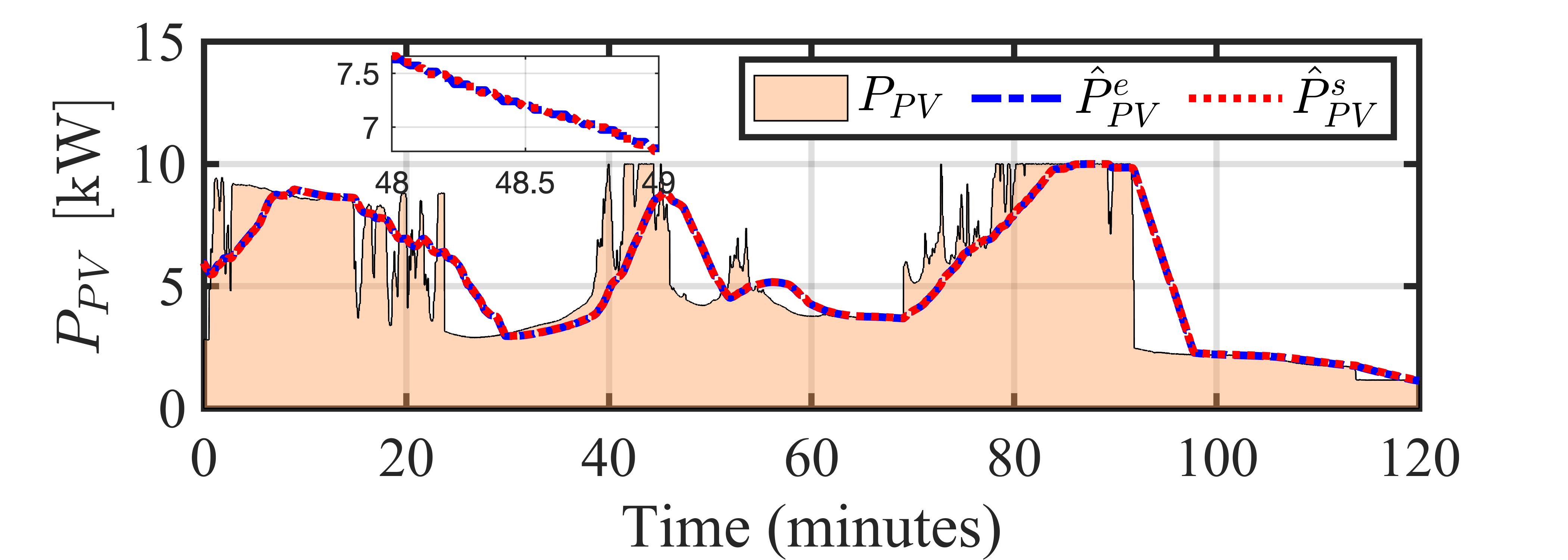}
        {(a)\par}
    \end{minipage}\hfill
    \begin{minipage}{0.45\textwidth}
        \centering
        \includegraphics[width=\linewidth]{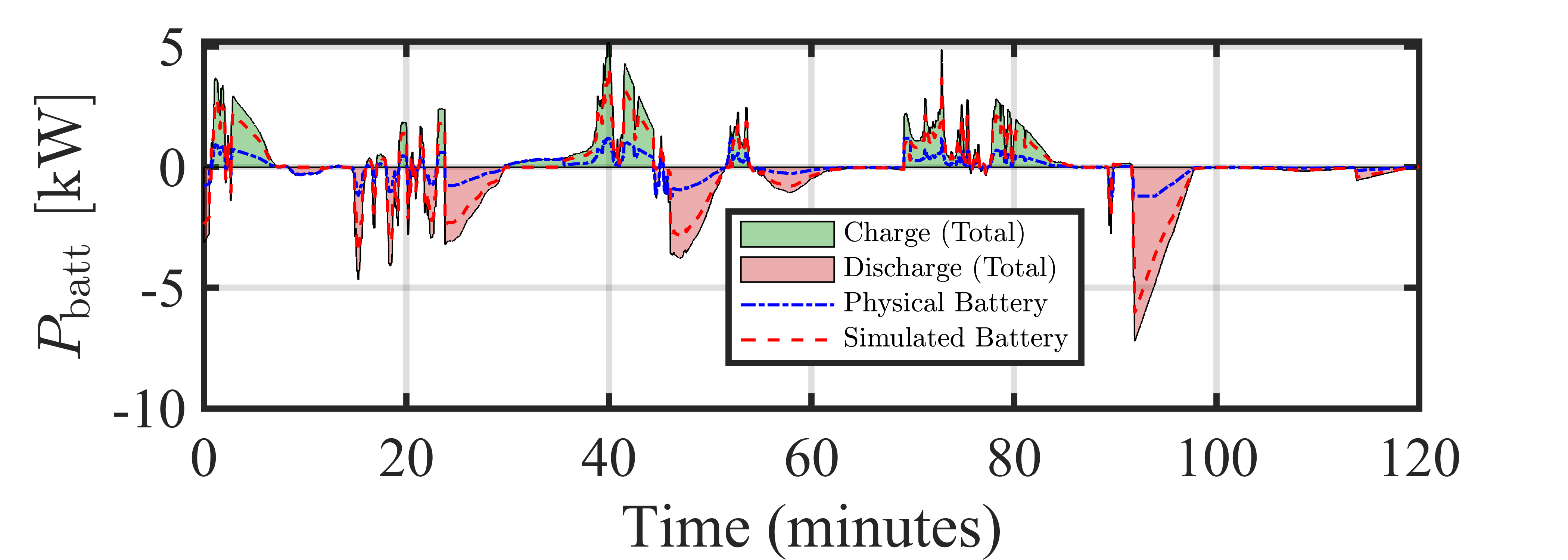}
        {(b)}
    \end{minipage}
     \begin{minipage}{0.45\textwidth}
        \centering
        \includegraphics[width=\linewidth]{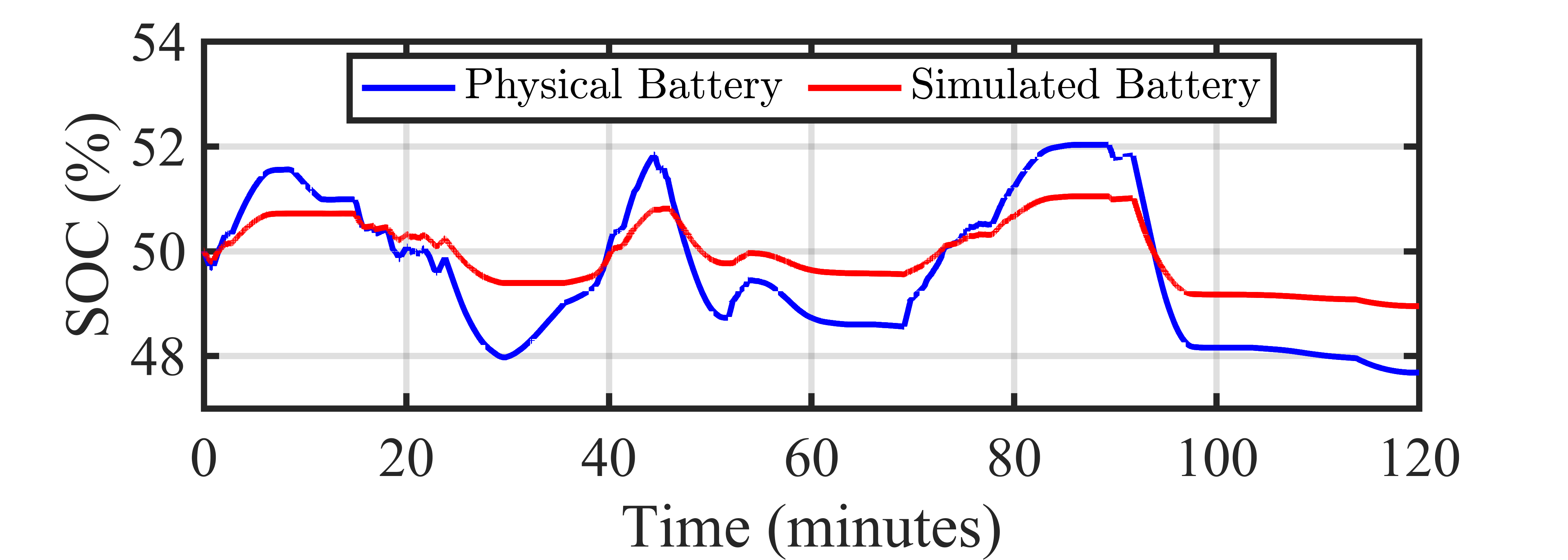}
        {(c)\par}
    \end{minipage}
     \begin{minipage}{0.45\textwidth}
        \centering
        \includegraphics[width=\linewidth]{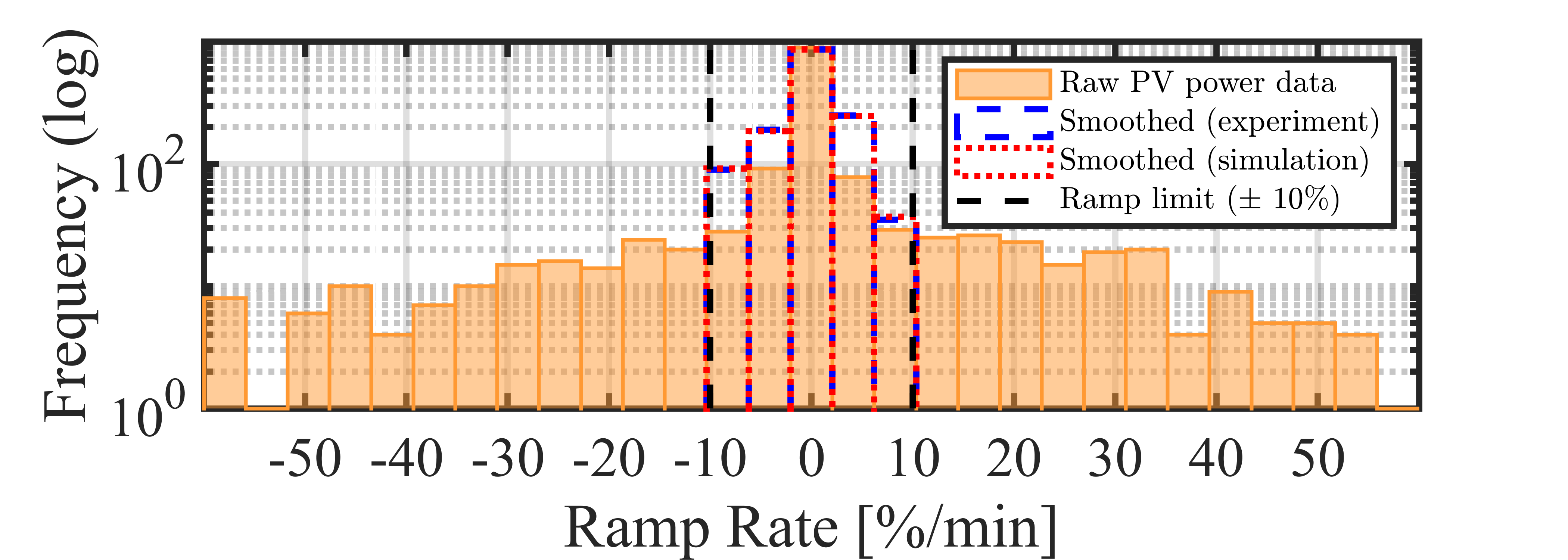}
        {(d)}
    \end{minipage}

    \caption{Two-hour snapshot of HIL experimental PV power smoothing using 5-second HSTC data and the ATL battery setup. (a) Raw and smoothed PV power, where $P_{\text{PV}}$ denotes the raw PV power, and $\hat{P}_{\text{PV}}^{e}$ and $\hat{P}_{\text{PV}}^{s}$ denote the experimentally measured and simulated smoothed PV power, respectively, (b) dispatched battery power of the physical and simulated batteries, (c) battery SOC, and (d) ramp-rate distributions.}
    \label{fig:PV_Smoothing}
\end{figure}

Fig.~\ref{fig:PV_Smoothing} presents a two-hour snapshot of the HIL experiment. The raw PV power exhibits a maximum ramp rate of 56~\%/min, significantly exceeding the $\pm10~\%$/min grid limit. Fig.\ref{fig:PV_Smoothing}(a) demonstrates that the BESS effectively reduces this to 9.45\%/min. The simulation and experimental results show close agreement, with a root mean square error (RMSE) of 0.25~kW (2.4~\% of maximum power). This error is attributed to communication latency, ADC quantization, sensor noise, and sampling effects. Fig.~\ref{fig:PV_Smoothing}(b) presents dispatched battery power for the physical and simulated batteries. Fig.~\ref{fig:PV_Smoothing}(c) shows the SOC of both batteries. The physical battery exhibits larger deviations due to its lower capacity and maximum power; however, both batteries remain within safe operating limits. Fig.~\ref{fig:PV_Smoothing}(d) illustrates the ramp rate distribution, showing reduced extreme events and zero grid-constraint violations. These results validate ATL's capability to implement and test PV smoothing strategies under multi-site operation.

\section{Conclusion}
\label{sec:conclusion}
This paper presented an experimental testbed for validating HES coordination and control strategies. The ATL platform integrates kW-scale PV, batteries, grid-tied inverters, and an electrolyzer with a real-time simulator. It enables modeling, control, and HIL validation of HES at both component and system levels. The solar smoothing experiment demonstrates ATL’s ability to test grid-service control strategies.

Future work will use the OPAL-RT real-time simulator to coordinate all subsystems within a unified control and communication framework targeting diverse operational objectives. Another direction will focus on implementing and validating GFL and GFM control strategies under weak-grid and transient conditions. These advancements will further establish the ATL as a comprehensive real-time facility for accelerated research and deployment of next-generation HES.

\section*{Acknowledgment}
The authors thank Peng Sang and Madiha Akbar at the University of Vermont for contributions to PV data streaming and to modeling and testing the electrolyzer in the ATL.
\bibliographystyle{IEEEtran}
\bibliography{refs.bib}

@techreport{Berkeley2025,
    author = {{Will Gorman} and {Joe Rand} and {Anna Cheyette} and {Joachim Seel},
{SeongeunJeong} and {Ryan Wiser}},
    title = {Hybrid Power Plants: Status of Operating and Proposed Plants} ,
    institution = {Lawrence Berkeley National Laboratory},
    year = {October 2025} 
}

@article{liang2016emerging,
  title={Emerging power quality challenges due to integration of renewable energy sources},
  author={Liang, Xiaodong},
  journal={IEEE Transactions on Industry Applications},
  year={2016}
}

@article{nassar2025role,
  title={The role of hybrid renewable energy systems in covering power shortages in public electricity grid: An economic, environmental and technical optimization analysis},
  author={Nassar, Yasser Fathi and El-Khozondar, Hala Jarallah and Fakher, Masoud Ali},
  journal={Journal of Energy Storage},
  year={2025},
  publisher={Elsevier}
}

@article{das2025research,
  title={Research Challenges and Opportunities of Utility-Scale Hybrid Power Plants},
  author={Das, Kaushik and Hansen, Anca D and others},
  journal={Wiley Interdisciplinary Reviews: Energy and Environment},
  year={2025},
  publisher={Wiley Online Library}
}

@inproceedings{stanev2020real,
  title={A Real Time Power Hardware in the Loop Test Bed for Power System Stability Studies},
  author={Stanev, Rad and Viglov, Kostadin and Nakov, Kamen and Asenov, Tsvetomir},
  booktitle={2020 12th Electrical Engineering Faculty Conference (BulEF)},
  year={2020}
}

@inproceedings{kroposki2012energy,
  title={Energy systems integration facilities at the national renewable energy laboratory},
  author={Kroposki, Benjamin and Mooney, David and Markel, Tony and Lundstrom, Blake},
  booktitle={2012 IEEE Energytech},
  year={2012}
}

@inproceedings{glover2012secure,
  title={Secure scalable microgrid test bed at sandia national laboratories},
  author={Glover, S and Neely, J and others},
  booktitle={2012 IEEE International Conference on Cyber Technology in Automation, Control, and Intelligent Systems},
  year={2012}
}

@article{turner2014design,
  title={Design and active control of a microgrid testbed},
  author={Turner, Greg and Kelley, Jay P and Storm, Caroline L and Wetz, David A and Lee, Wei-Jen},
  journal={IEEE Transactions on Smart Grid},
  year={2014}
}

@inproceedings{ghanbari2019hardware,
  title={Hardware-in-the-loop implementation of a grid connected pv system},
  author={Ghanbari, Niloofar and Bhattacharya, Subhashish},
  booktitle={2019 IEEE Industry Applications Society Annual Meeting},
  year={2019}
}

@article{mishra2024predictive,
  title={Predictive optimization of hybrid energy systems with temperature dependency},
  author={Mishra, Tanmay and Pandey, Amritanshu and Almassalkhi, Mads R},
  journal={Electric Power Systems Research},
  year={2024},
  publisher={Elsevier}
}

@inproceedings{omidi2025towards,
  title={Towards input-convex neural network modeling for battery optimization in power systems},
  author={Omidi, Arash and Mishra, Tanmay and Almassalkhi, Mads R},
  booktitle={2025 American Control Conference (ACC)},
  year={2025}
}

@article{ayubirad2025modeling,
  title={Modeling and Constraint-Aware Control of Pressure Dynamics in Water Electrolysis Systems},
  author={Ayubirad, Mostafaali and Akbar, Madiha and Ossareh, Hamid R},
  journal={arXiv preprint arXiv:2505.16935},
  year={2025}
}

@inproceedings{youssef2015dds,
  title={DDS based interoperability framework for smart grid testbed infrastructure},
  author={Youssef, Tarek A and Elsayed, Ahmed T and Mohammed, Osama A},
  booktitle={2015 IEEE 15th International Conference on Environment and Electrical Engineering (EEEIC)},
  year={2015}
}

@misc{UNIFI2025,
title = "UNIFI's Grid-Forming (GFM) Inverter Reference Design: A Tutorial on Modeling, Control, and Experimental Implementation of GFM Inverters",
author = "Debjyoti Chatterjee and Jakob Triemstra and others",
year = "2025",
doi = "10.2172/2583465",
language = "American English",
type = "Other",
}

@article{sang2024circuit,
  title={Circuit-theoretic joint parameter-state estimation of utility-scale photovoltaic, battery, and grid systems},
  author={Sang, Peng and Pandey, Amritanshu},
  journal={Sustainable Energy, Grids and Networks},
  year={2025},
  publisher={Elsevier}
}

@article{essakiappan2021multi,
  title={A multi-site networked hardware-in-the-loop platform for evaluation of interoperability and distributed intelligence at grid-edge},
  author={Essakiappan, Somasundaram and Chowdhury, Prithwiraj Roy and others},
  journal={IEEE Open Access Journal of Power and Energy},
  year={2021},
  publisher={IEEE}
}

@article{martins2019comparative,
  title={Comparative study of ramp-rate control algorithms for PV with energy storage systems},
  author={Martins, Jo{\~a}o and Spataru, Sergiu and Sera, Dezso and Stroe, Daniel-Ioan and Lashab, Abderezak},
  journal={Energies},
  year={2019},
  publisher={MDPI}
}

\end{document}